\documentclass[preprint2]{aastex}

\shorttitle{Causalit\'e revisit\'ee et relativit\'e g\'en\'erale}
\shortauthors{Bois et al.}

\begin{document}


\title{
    L'impact de la structure chrono-g\'eom\'etrique
    de l'espace-temps sur la causalit\'e}

\author{E. Bois\altaffilmark{1,2}}
\email{Eric.Bois@obs.u-bordeaux1.fr}
 \and \author{E. Trelut\altaffilmark{1,2}}
\email{e.trelut@episteme.u-bordeaux.fr}


\altaffiltext{1}{Observatoire Aquitain des Sciences de l'Univers,
UMR CNRS/INSU 5804, B.P.89, F-33270 Floirac, France.}
\altaffiltext{2}{Laboratoire \textit{\'EPIST\'EM\'E}, Universit\'e
Bordeaux 1, 40, rue Lamartine, F-33400 Talence, France.}


\begin{abstract}
{L'article porte sur une analyse \'epist\'emologique de la
structure courante de la causalit\'e physique face \`a
l'\'evolution des concepts de temps et d'espace de la m\'ecanique
classique \`a la relativit\'e g\'en\'erale. Le sens usuel de la
causalit\'e se caract\'erise par la contrainte de la s\'eparation
temporelle entre la cause et l'effet selon une succession
d'instants ordonn\'es lin\'eairement dans l'ensemble des r\'eels.
Or, d'une part, la racine conceptuelle du param\`etre appel\'e
``temps'' (hom\'eomorphe \`a la droite des r\'eels) est celle du
temps extrins\`eque et absolu de Newton. Et, d'autre part, la
relativit\'e g\'en\'erale indique que la coordonn\'ee de genre
temps se pr\'esente comme une dimension non-extrins\`eque \`a la
mati\`ere et de surcro\^\i t dynamiquement m\^el\'ee aux
coordonn\'ees de genre espace. L'espace-temps courbe de la
relativit\'e g\'en\'erale relie en effet le tenseur m\'etrique
$g_{\mu\nu}$ aux distributions d'\'energie d\'ecrites par les
\'equations du champ d'Einstein. Dans cette perspective, la
logique de l'ant\'eriorit\'e causale ne peut plus \^etre r\'eduite
\`a une r\'ef\'erence de type chronologique. Le pr\'esent article
conduit \`a l'\'elaboration conceptuelle d'une contrainte
d'ant\'eriorit\'e spatio-temporelle intrins\`equement li\'ee \`a
la mati\`ere, c'est-\`a-dire une notion causale dans laquelle il
est n\'ecessaire de distinguer \textit{ant\'eriorit\'e strictement
temporelle} et \textit{causalit\'e}.}
\end{abstract}


\section{Introduction: causalit\'e et temporalit\'e}

\hfill "All concepts even those which are closest to experience,
are from the point of view of logic freely chosen conventions,
\hfill just as is the case with the concept of causality" \hfill
A. Einstein

\subsection{Pr\'eliminaire}

Selon l'acception commune, une relation causale est d\'{e}finie
entre deux \'{e}v\'{e}nements de telle sorte que la production du
second par le premier implique un ordre temporel
d\'{e}termin\'{e}\,\footnote{Selon le linguiste B. Pottier,
``~L'exp\'erience du temps a toujours s\'eduit le philosophe, le
psychologue et le linguiste. La part de la conceptualisation
culturalis\'{e}e est certes importante, mais le temps aura
toujours deux caract\'eristiques in\'eluctables~ : il est
naturellement irr\'eversible (il le devient par l'imaginaire), il
s'impose \`{a} l'homme. L'homme subit le temps, alors qu'il peut
dominer l'espace. Le temps t$_{{\rm 0}{\rm} {\rm} }$ est
consubstantiel \`{a} la pens\'{e}e. L'avant et l'apr\`{e}s ne
peuvent \^etre vus qu'\`{a} partir de t$_{{\rm 0}}$. ``~O\`{u}
\^{e}tes vous~?~'' a du sens, ``~Quand \^{e}tes-vous~?~'' n'en a
pas, \`{a} moins d'entrer dans la fiction~''. Et de m\^{e}me, le
champ d'application notionnelle s'organise \`{a} partir de
l'\textit{ego}~: ``~Le rep\`{e}re de l'\textit{ego-t}$_{0{\rm} }$
permet de consid\'erer l'avant et l'apr\`{e}s~et d'y fixer des
rep\`{e}res secondaires en nombre non-limit\'{e}, pour des
constructions aspectotemporelles. Une application notionnelle de
cette vision est la source des relations logicos\'emantiques comme
(``~s'il pleut, alors je ne sors pas~; je ne suis pas sorti,
puisqu'il pleuvait~'').}. Autrement dit, on range habituellement
des faits quelconques dans un ordre causal qui est aussi l'ordre
chronologique de leur succession. Cet usage commun rappelle
l'effort humien de r\'eduction de la relation causale \`a une
simple succession temporelle d'\'{e}v\'{e}nements empiriques~:
\begin{quote}
``~Nous trouvons, affirme Hume, seulement que l'un suit l'autre
effectivement, en fait~''.
\end{quote}
Follon (1998) rapporte une synth\`ese pertinente de l'histoire de
la causalit\'e \`a travers ses diff\'erentes conceptions et ses
grands d\'ebats (pythagoriciens, Platon, Aristote, Bacon,
Descartes, Spinoza, Leibniz, Hume...). Des travaux r\'{e}cents sur
la causalit\'{e} (Salmon 1970, 1984) et
Suppes (1970), par exemple) poursuivent la perspective
empiriste de Hume, mais dans un cadre logique o\`u l'ordre causal
est r\'eductible \`a l'ordre temporel; d'autres, au contraire,
suite \`a la relativit\'e de l'ordre temporel depuis la
d\'ecouverte de la loi r\'egissant la propagation de la lumi\`ere,
cherchent comme Robb (1914), Carnap (1925) et Reichenbach (1922),
ou plus r\'ecemment Papineau (1985, 1986, 1989, 1991, 1993) \`a
construire par voie axiomatique l'ordre temporel par l'ordre
causal; enfin, certains travaux argumentent en faveur de
l'irr\'eductibilit\'e de la relation causale \`a l'asym\'etrie
temporelle (Cartwright 1979, 1983, 1989; Miller 1987; Kistler 1999;
Woodward 1992). Notre analyse s'inscrit \`a la suite des travaux de
cette derni\`ere s\'erie d'auteurs. Cependant, il ne s'agit pas
pour nous de peser le pour et le contre des arguments favorables
\`a la th\`ese logique d'une th\'eorie causale du temps. Notre
d\'emarche, analogue en ce sens \`a celle de Heller (1990, 1991)
s'enracine dans les propri\'et\'es topologiques des espaces de
Riemann qui renouvellent le concept d'ant\'eriorit\'e temporelle.

\subsection{La notion de temps en physique classique}

Avec Newton, la causalit\'e ne consiste pas \`a enregistrer des
rapports de succession au cours du temps. Elle est une
d\'etermination fonctionnelle d'ant\'ec\'edents \`a cons\'equents,
d'un ph\'enom\`ene \`a un autre, d\'ecrite par l'\'equation
fondamentale de la m\'ecanique rationnelle
\begin{equation}\label{1}
  \mathbf{F} = m \mathbf{a},
\end{equation}
qui \'etablit une relation \textit{analytique} entre la force
$\mathbf{F}$ et l'acc\'el\'eration $\mathbf{a}$, avec

\begin{equation}\label{2}
  \frac{{d^2 \mathbf{x}}}{{dt^2 }} = \mathbf{a}
\end{equation}
o\`u $\mathbf{x}$ est le vecteur position d'une masse ponctuelle
$m$ \`a un instant de temps $t$. La force et la trajectoire,
c'est-\`a-dire la cause et l'effet, sont reli\'ees par une
relation diff\'erentielle dans laquelle est inscrite un
\'el\'ement infinit\'esimal de temps. D\`es lors, le temps $t$ des
\'equations de la m\'ecanique newtonienne est un temps
construit\,\footnote{Concernant la conception classique du temps,
M. Paty fait remarquer que l'on ``~a pas pris assez garde que le
temps newtonien n'est pas tant donn\'{e} que construit, m\^{e}me
s'il n'est pr\'{e}sent\'{e} comme tel et que la sp\'ecification du
``temps absolu, vrai et math\'ematique'' a, en r\'{e}alit\'{e},
surtout le r\^ole de pr\'{e}parer la condition d'une formulation
plus radicale du concept de temps, sous les esp\`{e}ces d'une
grandeur math\'ematis\'{e}e, singuli\`{e}re et \`{a} variation
continue, c'est-\`a-dire diff\'erentielle~''.},
math\'{e}matis\'{e} sous la forme d'un \textit{param\`etre
d'\'evolution}, dont la variation continue parcourt l'ensemble
ordonn\'e des r\'eels $(\mathbb{R}, \leq )$. La math\'ematisation
du temps consiste donc \`{a} faire correspondre \`{a} chaque point
d'une vari\'{e}t\'{e} unidimensionnelle un nombre r\'{e}el
d\'{e}termin\'{e}; inversement \`{a} tout nombre r\'{e}el
correspond un point de cet espace et un seul. En outre, le point
origine d\'epend d'un choix arbitraire sur la vari\'{e}t\'{e}
unidimensionnelle.

Une autre forme du temps, \`a savoir le temps comme coordonn\'ee
ou dimension, est le fait que les ph\'enom\`enes physiques
continus peuvent \^etre repr\'esent\'es par des champs ob\'eissant
\`a des \'equations aux d\'eriv\'ees partielles (A. Lautman 1946).
En particulier, dans le cadre de la
th\'eorie relativiste du champ \'electromagn\'etique, les
conditions spatio-temporelles (principe de relativit\'{e}
galil\'eenne et contrainte de l'invariance de la vitesse de
propagation des actions) introduisent des relations structurelles
de possibilit\'e de connexion entre les \'ev\`enements. De la
sorte, la causalit\'e semble soumise \`{a} la contrainte physique
d'ant\'{e}riorit\'{e} temporelle\footnote{Dans la physique
newtonienne les relations causales entre des points s\'epar\'es
temporellement sont des faits d'ordre empiriques ; mais, souligne
Gr\"unbaum (1968), ``~[...]in the context of the STR, the
presumed facts of clock behavior under transport destroy the
physical foundation on which the \textit{empirical} character of
the question ``are \textit{all} pairs of causally connectible
events time-separated?'' depends. And thus causal connectibility
becomes \textit{constitutive} of \textit{absolute time
separation}, while causal non-connectibility becomes constitutive
of invariant \textit{topological} simultaneity. This
constitutivity of absolute time order by causal connectibility and
non connectibility is \textit{presupposed} by the light-signal
method of synchronizing clocks and hence by the time relations
which ensue from the time numbers thus assigned by clocks''.}.
Cela signifie-t-il que la conception relativiste est, comme le
fait remarquer de Broglie (1974, pp. 88-89),
\begin{quote}
``[...] en quelque sorte le couronnement de la physique
classique'',
\end{quote}
\noindent qui associerait la d\'etermination newtonienne \`a
l'ant\'eriorit\'e temporelle de Hume?

\subsection{Probl\'ematique}

La notion de temps coordonn\'ee, ou dimension orient\'ee, est
formellement traduite par l'existence d'une diff\'erence de signe
dans les termes composant la forme quadratique fondamentale $ds^2$
d'une vari\'et\'e espace-temps $\mathcal{M}$ \`a quatre dimensions
(celle-ci est qualifi\'ee improprement euclidienne ou improprement
riemannienne selon que la courbure des g\'eod\'esiques est nulle
ou non). Pour une vari\'et\'e $\mathcal{M}$ pseudo-euclidienne -
l'espace-temps de Minkowski $(\mathcal{M}, \eta)$ - la m\'etrique
$ \eta_{\mu\nu}$ refl\`ete d'embl\'ee la structure globale de la
vari\'et\'e (Auyang 1995, pp. 38-39 et Friedman 1983, p.
37). Au contraire, pour une vari\'et\'e $\mathcal{M}$
pseudo-riemannienne, la m\'etrique $ g_{\mu\nu}$ pr\'esente
partout une signature de Lorentz sans fixer la structure globale
de la vari\'et\'e espace-temps. Pour une m\'etrique donn\'ee, il
existe en effet des espace-temps topologiquement distincts (Lachieze-Rey 1995).
Par exemple, \'etant donn\'ee la
vari\'et\'e $\mathcal{M}_{4}$ hom\'eomorphe au produit topologique
de la vari\'et\'e $\mathcal{M}_{3}$ (l'espace physique) et de la
droite r\'eelle $\mathbb{R}$ (le temps), l'on montre que la
vari\'et\'e $\mathcal{M}_{3}$ \`a courbure constante peut \^etre
hom\'eomorphe \`a une infinit\'e de produits topologiques
(lesquels sont class\'es en huit classes d'espaces
tridimensionnels) (Lachieze-Rey 1995).

S'il existe bien une pluralit\'e de topologies associ\'ees \`a
l'espace, l'en\-sem\-ble des instants $(T, \leq)$ forme un espace
topologique unidimensionnel hom\'eomorphe au seul ensemble
$\mathbb{R}$ des nombres r\'eels, ce qui exclut la topologie du
cercle construite \`a partir de l'identification des instants \`a
l'infini dans le temps. On sait en effet (Delachet 1974, \S\S 14
et 29) que l'ensemble $\mathbb{R}$ des nombres r\'eels est
identifiable \`a la droite euclidienne. La droite num\'erique
achev\'ee $\mathbb{\overline{R}}$, c'est-\`a-dire l'ensemble $
\mathbb{R} \cup \{ - \infty , + \infty \}$, est totalement
ordonn\'ee par la relation d'ordre totale
$\leq$.\footnote{L'ensemble $N$ des intervalles ouverts et
demi-droites ouvertes de $\mathbb{\overline{R}}$ d\'efinit une
topologie sur $(\mathbb{\overline{R}}, \leq)$, appel\'ee topologie
d'ordre ou encore \textit{topologie usuelle} de
$(\mathbb{\overline{R}}, \leq)$.}

On est donc conduit \`a admettre que l'ant\'eriorit\'e de la cause
sur l'effet et la topologie du temps lin\'eaire sont fortement
imbriqu\'ees en ce sens que la {\it topologie du temps} se trouve
d\'ej\`a dans les hypoth\`eses sur la {\it causalit\'e
usuelle}:\footnote{On conviendra de l'usage de l'expression
"topologie du temps" pour d\'esigner l'espace topologique
unidimensionnel hom\'eomorphe \`a l'ensemble $\mathbb{R}$ des
nombres r\'eels sur lequel est construite la coordonn\'ee de genre
temps.}
\begin{itemize}
\item \textbf{Axiome 1}. La {\it causalit\'e usuelle} contient
implicitement une topologie du temps; \item \textbf{Axiome 2}. La
cause usuelle est, \textit{a priori}, toujours ant\'erieure dans
le temps \`a l'effet; \item \textbf{Axiome 3}. La {\it causalit\'e
usuelle} contraint la topologie du temps \`a s'identifier \`a
celle de la droite euclidienne.
\end{itemize}
Ces trois axiomes d\'efinissent ainsi la structure de la
causalit\'e usuelle commun\'ement admise en sciences de la nature.
Toutefois, selon la th\'eorie de la relativit\'e g\'en\'erale, les
coefficients $g_{\mu\nu}$ de la m\'etrique d\'ependent en chaque
point de la distribution de mati\`ere-\'energie.  D\'esormais, le
temps construit est non seulement alg\'ebriquement m\^el\'e \`a
l'espace,\footnote{Comme le dit Minkowski (1908): ``La vision de l'espace
et du temps que je souhaite vous exposer a germ\'e sur le sol de
la physique exp\'erimentale d'o\`u elle puise sa force. Elle est
radicale. D\'esormais l'espace en soi, le temps en soi, sont
d\'eclar\'es rel\'egu\'es au royaume des ombres, et seule une
sorte d'union des deux pourra pr\'eserver une r\'ealit\'e
ind\'ependante''.} mais il est aussi et surtout, avec
l'espace, une ``qualit\'e structurale'' ins\'eparable de la
mati\`ere. Il en r\'esulte que les quatre nombres r\'eels
coordonn\'es arbitrairement \`a chaque point dans $(\mathcal{M},
g_{\mu\nu})$ ne peuvent signifier des grandeurs physiques de temps
et d'espace ayant un quelconque rapport conceptuel avec ces
m\^emes grandeurs \`a l'\oe uvre en causalit\'e usuelle. L'\'ecart
n'est pas r\'eductible \`a la seule diff\'erence axiomatique et
"technique" de construction du temps et de l'espace, il est
conceptuel. Par cons\'equent, des coordonn\'ees de genre temps et
espace ainsi construites en relativit\'e g\'en\'erale, l'on est
conduit \`a une conception du temps et de l'espace qui n'ont plus
la capacit\'e d'exister par eux-m\^emes de mani\`ere
ind\'ependante et s\'epar\'ee.\footnote{Comme le dit Einstein (1983):
``Soit donn\'e, par exemple, un champ de gravitation pur d\'ecrit
par les $g_{ik}$ (comme fonction des coordonn\'ees) en r\'esolvant
les \'equations de la gravitation. Si l'on suppose le champ de
gravitation, c'est-\`a-dire les $g_{ik}$, \'elimin\'e, il ne reste
pas un espace du type (1) [espace-temps de Minkowski], mais
absolument \textit{rien}, pas m\^eme un "espace
topologique".}

L'\'equation d'Einstein s'\'ecrit:

\begin{equation}\label{3}
  G_{\mu\nu} + \Lambda g_{\mu\nu} =  - \kappa T_{\mu\nu}
\end{equation}
o\`u $G_{\mu\nu}$ est le tenseur d'Einstein, $g_{\mu\nu}$ la
m\'etrique de l'espace-temps, $T_{\mu\nu}$ le tenseur
impulsion-\'energie, $\Lambda$ la constante cosmologique et
$\kappa$ le c\oe fficient de proportionalit\'e \'egal \`a
$\kappa=8\pi G/c^{4}$ ($G$ \'etant la constante fondamentale de la
gravitation, $c$ celle de la vitesse de la lumi\`ere
interpr\'et\'ee comme la constante de structure de
l'espace-temps). Il s'av\`ere que l'existence d'un axe temporel
n'est pas une {\it propri\'et\'e g\'en\'erique} (Heller 1996) de
l'ensemble des solutions possibles de
(3).\footnote{Par exemple, G\"{o}del (1949) a propos\'e une
solution d'un univers en rotation n'admettant pas d'axe temporel
universel, les courbes de genre temps \'etant ferm\'ees.} La
vari\'et\'e espace-temps $(\mathcal{M}, g_{\mu\nu})$ doit \^{e}tre
impl\'{e}ment\'{e}e d'une hi\'erarchie de conditions de
causalit\'{e} (Carter 1971) portant sur les
propri\'{e}t\'{e}s globales. En d'autres termes, afin de
pr\'eserver la {\it causalit\'e usuelle}, entendons ici l'axiome
3, on associe la {\it topologie du temps} aux propri\'et\'es
globales de certaines classes d'espace-temps que l'on qualifie
d'admissibles. Mais quelle preuve concluante d'impossibilit\'e
nous permet de trier des possibles topologiques et de dire que les
mod\`eles d'espace-temps temporellement "anormaux" n'existent pas?
Le seul argument donn\'e invoque la causalit\'e usuelle. En effet,
dans de tels espace-temps (anormaux), un \'ev\'enement pourrait
avoir une influence causale sur son propre pass\'e (Hawking 1974, \S6.4);
ce qui n'est pas admissible. Pr\'ecis\'ement, l'adjectif ``admissible''
est utilis\'e pour
signifier que la restriction de la causalit\'e \`a
l'ant\'eriorit\'e temporelle contraint le couplage g\'eom\'etrie
mati\`ere-\'energie \`a ne pas consid\'erer comme physiquement
raisonnables, via les tenseurs d'Einstein et
d'impulsion-\'energie, des espace-temps temporellement anormaux,
en d\'epit du fait qu'ils sont th\'eoriquement permis. En fait,
l'impossibilit\'e (ou l'inadmissibilit\'e) d'anomalies causales
n'est pas issue du syst\`eme des \'equations einsteiniennes du
champ; elle n'est donc pas dynamique, mais d'ordre
\'epist\'emique.\footnote{Selon A.N. Whitehead (1968), la
pens\'ee occidentale ``[...] has been hampered by the tacit
presupposition of the necessity of a static spatio-temporal, and
physical forms of order. [...] In current literature we find the
same authors denying infractions of natural order, and denying any
reason for such denial, and denying any justification for a
philosophical search for reasons justifying their own
denials''(lecture five).  Il en r\'esulte ``[...] a complete
muddle in scientific thought, in philosophic cosmology, and in
epistemology. But any doctrine which does not implicitly
presuppose this point of view is assailed as unintelligible''
(lecture seven), (voir aussi Sklar 1985).}

\subsection{Remarques sur la m\'ethode}
Nous voudrions souligner la position usuelle qui fait usage d'une
causalit\'e avec r\'ef\'erence au temps newtonien. Les
propri\'et\'es newtoniennes du temps sont les suivantes:
universalit\'e (le temps a les m\^emes propri\'et\'es dans tout
l'univers), absolu, mesurable, synchronisable, \'eternel, distinct
de l'espace, uni-dimensionnel (topologie de la droite infinie),
orientable, irr\'eversible et conforme au d\'eterminisme. De la
sorte, la causalit\'e usuelle interdit les boucles de temps.
Certes, ce temps newtonien est critiquable, \`a savoir qu'il
n\'ecessite un cadre ind\'eformable; il semble \'egalement manquer
de consistance et de r\'ealit\'e. N\'eanmoins, les propri\'et\'es
du temps en relativit\'e sont les suivantes: vitesse finie de
propagation des ondes \'electromagn\'etiques, absence de
simultan\'eit\'e absolue, non-absolu (le temps et l'espace sont
associ\'es et m\^el\'es dans la structure espace-temps) et
\'elastique (en tant qu'il d\'epend du contenu mati\`ere-\'energie
de l'univers en relativit\'e g\'en\'erale). En ce sens, \textit{a
priori}, la causalit\'e et l'orientabilit\'e du temps ne sont plus
assur\'es globalement. Une synchronisation universelle n'est pas
toujours possible, et la simultan\'eit\'e  absolue n'existe plus.
Les propri\'et\'es du temps newtonien sont donc perdues.
Cependant, g\'en\'eralement on construit des mod\`eles
cosmologiques relativistes (les mod\`eles de type
Friedman-Lema\^\i tre-Robertson-Walker (FLRW), par exemple), de
telle sorte que l'on puisse retrouver certaines propri\'et\'es
newtoniennes du temps, et pr\'eserver ainsi la causalit\'e au sens
usuel (voir section 3). Or par ce pr\'esent travail, plut\^ot que
de retrouver par r\'eduction des concepts de la relativit\'e
g\'en\'erale les propri\'et\'es du temps newtonien, nous
choisissons d'enrichir la causalit\'e elle-m\^eme de ce
qu'implique les propri\'et\'es du temps relativiste.

En section 2 de cet article, nous pr\'esentons une br\`eve
description de la structure causale induite par la validit\'e
locale de la m\'etrique de Minkowski $(\eta)$ pour une vari\'et\'e
espace-temps $\mathcal{M}$ munie d'une m\'etrique $(g)$ de
signature hyperbolique $(1, 3)$. Dans un article r\'ecent, l'un de
nous (Bois 2000) pr\'esente quelques n\'ecessit\'es de revisiter
les \'el\'ements de la causalit\'e, notamment \`a partir des
implications conceptuelles de la relativit\'e g\'en\'erale
(couplage dynamique g\'eom\'etrie-mati\`ere). Dans cette
perspective, nous reprenons en section 3 la construction
topologique en feuilletage d'une vari\'et\'e espace-temps
$\mathcal{M}$ en une m\'etrique d'espace associ\'ee \`a une
direction de temps et levons la contrainte d'ant\'eriorit\'e
strictement temporelle. Enfin, la section 4 conclut en proposant
une classification de l'ant\'eriorit\'e causale en trois niveaux.

\section{Causalit\'e locale et c\^one isotrope}

A tout construit riemannien $\mathcal{M}$ de dimensions 4, il est
toujours possible de choisir, en un point-\'ev\'enement $p$
quelconque de $\mathcal{M}$, un syst\`eme de coordonn\'ees local
d\'efini sur le plan tangent en $p$, $T_p (\mathcal{M})$, tel que:
\begin{enumerate}
  \item $\mathcal{M}$ poss\`ede localement la m\^eme structure
  m\'etrique d'espace-temps qu'en relativit\'e restreinte,
  c'est-\`a-dire le tenseur m\'etrique $ g_{\mu\nu}$ se r\'eduise
  \`a la forme pseudo-euclidienne $ \eta_{\mu\nu}$,\footnote{O\`u
  $\eta_{\mu\nu}=diag(1, - 1, - 1, - 1)$
  est la matrice de Minkowski de signature $(-2)$.}
  \item Les d\'eriv\'ees premi\`eres du tenseur m\'etrique soient
  localement
  nulles, c'est-\`a-dire $\Gamma _{\nu \lambda}^\mu  = 0$.
\end{enumerate}
De ce fait, la structure m\'etrique de l'espace-temps $T_p
(\mathcal{M})$, tangent en $p$ \`a $\mathcal{M}$, est
caract\'eris\'e par l'invariance dans les changements de
syst\`emes de coordonn\'ees galil\'eens de l'intervalle
\'el\'ementaire $ds^{2}$ d'\'equation:
\begin{equation}\label{}
  ds^2  = dt^2  - dx^2  - dy^2  - dz^2
\end{equation}
par les transformations lin\'eaires du groupe de Lorentz
$SO(1,3)$.\,\footnote{Pour la suite on \'ecrira $ \{ t,x,y,z\}  =
\{ x^0 ,x^1 ,x^2 ,x^3 \} = \{ x^\mu  \} _{\mu  = 0,1,2,3}$.}
\begin{quote}
``~Par ce proc\'{e}d\'{e} le temps perdit son caract\`ere absolu
et fut adjoint aux coordonn\'ees spatiales comme une grandeur
ayant presque le m\^{e}me type alg\'{e}brique. Le caract\`{e}re
absolu du temps et particuli\`erement celui de la
simultan\'{e}it\'{e} \'{e}tait d\'etruit et la description
quadridimensionnelle fut introduite comme la seule qui fut
ad\'equate~''.\footnote{(Einstein, 1936)}
\end{quote}

Ces transformations de Lorentz formalisent la nouveaut\'e
conceptuelle de l'espace-temps con\c cu d\'esormais comme une
structure "m\^elant" alg\'ebri\-quement la construction des
coordonn\'ees de temps et d'espace.

Par suite, deux sortes d'objets g\'eom\'etriques sont invariants,
\`a savoir le c\^one isotrope d'\'equation $ds^{2}= 0$ et une
classe de deux familles de droites, les droites de genre temps
(situ\'ees \`a l'int\'erieur du c\^one) et les droites de genre
espace (\`a l'ext\'erieur du c\^one).

Le c\^one isotrope et la classe des deux familles de droites qui
partagent $ \mathcal{M}$ en trois r\'egions: le futur temporel,
\`a l'int\'erieur de la nappe sup\'erieure du c\^one isotrope
$\Gamma^ + $; le pass\'e temporel, \`a l'int\'erieur de la nappe
inf\'erieure du c\^one $ \Gamma^ - $; et l'ailleurs (spatial)
$I^{0} $, \`a l'ext\'erieur du m\^eme c\^one. En effet, la
diff\'erence de signe dans les termes de la somme des carr\'es de
la forme quadratique $ds^{2} $ fait exister un temps coordonn\'e
et orient\'e. Les droites isotropes ($ ds^2  = 0 $) issues d'un
point quelconque de $ \mathcal{M}$ d\'efinissent un c\^one
isotrope de l'espace-temps attach\'e en ce point et toutes les
directions situ\'ees \`a l'int\'erieur de ce c\^one satisfont
l'in\'egalit\'e $ ds^2 > 0$.

Il s'ensuit \'egalement qu'un point quelconque $q$ de $
\mathcal{M}$ situ\'e dans $\Gamma^{+}$ du c\^one isotrope
attach\'e en un point $p$ est dans le futur de ce point, soit $ q
\in I^ +  (p) $. La pseudo-longueur du quadri-vecteur $\mathbf{u}$
qui joint les points-\'ev\'enements $p$ et $q$ \'etant positive,
$\mathbf{u}$ est de genre temps et orient\'e vers le futur. Ainsi,
pour tous les points $ q \in I^ + (p) $, $t - t_{0} = t > 0$,
$(t_{0} = 0$ est le temps au point $p$), et cela quel que soit le
syst\`eme de coordonn\'ees galil\'een.

Supposons une ligne d'univers $C$ quelconque d'une particule
anim\'ee d'un mouvement rectiligne uniforme par rapport \`a un
r\'ef\'erentiel galil\'een $(R)$, d\'efinie par la donn\'ee des
points $x^{\mu}$, o\`u:
\begin{equation}\label{}
  x^\mu = x^\mu (\xi ) = (x^0 (\xi ),x^1 (\xi ),x^2 (\xi ),x^3 (\xi))
\end{equation}
en fonction d'un param\`etre r\'eel arbitraire $\xi $. La
trajectoire spatio-temporelle de la particule est une fonction
monotone et orient\'ee vers les $\xi $ croissants. D\`es lors on
conclut qu'un quadri-vecteur vitesse $ \mathbf{u}$, tangent \`a sa
ligne d'univers $C$ d\'ecrite par $ x^\mu   = x^\mu (\xi ) $, est:

\begin{itemize}
  \item de genre temps et orient\'e vers le futur si:
\begin{equation}\label{}
    \eta _{\mu \nu } ({{dx^\mu  } \mathord{\left/
 {\vphantom {{dx^\mu  } {d\xi }}} \right.
 \kern-\nulldelimiterspace} {d\xi }})({{dx^\nu  } \mathord{\left/
 {\vphantom {{dx^\nu  } {d\xi }}} \right.
 \kern-\nulldelimiterspace} {d\xi }}) = k > 0,
\end{equation}

  \item et isotrope si:
  \begin{equation}\label{}
    k = 0.
\end{equation}
\end{itemize}

Associons \`a chaque point $p$ de $C(\xi)$ un rep\`ere naturel de
coordonn\'ees lorentziennes $(R')$, de vitesse $ \mathbf{u}$ par
rapport \`a $(R)$, auquel se trouve li\'ee une horloge id\'eale.
En choisissant un param\'etrage d\'efini par l'arc de
pseudo-longueur $s$ de la trajectoire $C$ dont l'\'el\'ement
infini\'esimal est $ds^{2}$, il s'ensuit que le temps propre
$d\tau$ mesur\'e par l'horloge attach\'ee au r\'ef\'erentiel
$(R')$ est toujours inf\'erieur au temps $ dt$ du r\'ef\'erentiel
$(R)$:
\begin{equation}\label{}
    d\tau ^2  = dt^2 \left( {1 - \frac{{\left\| \mathbf{u} \right\|^2
}}{{c^2 }}} \right)^{\frac{1}{2}}  < dt^2 .
\end{equation}

Consid\'erons \`a pr\'esent la pseudo-longueur de l'arc de courbe
$s$ joignant deux points quelconques $p$ et $q$ de $C(\tau)$ dont
la mesure est donn\'ee par l'int\'egrale suivante:
\begin{equation}\label{}
    \int {ds = \int_p^q {\left( {\eta _{\mu \nu } \frac{{dx^\mu
}}{{d\tau }}\frac{{dx^\nu  }}{{d\tau }}} \right)} }
^{{\raise0.5ex\hbox{$\scriptstyle 1$} \kern-0.1em/\kern-0.15em
\lower0.25ex\hbox{$\scriptstyle 2$}}} d\tau .
\end{equation}

Sa valeur est maximale si ces deux points $p$ et $q$ sont reli\'es
par les relations d'ant\'eriorit\'e satisfaisant aux crit\`eres de
Kronheimer et Penrose (1967) pour un espace-temps
\textit{causal} :

\begin{enumerate}
  \item La relation d'ant\'eriorit\'e
`horismotique',\footnote{L'expression grecque horismos signifie
horizon ou limite.} (AH) (``horismotic precedence'')
d\'etermin\'ee par les droites isotropes d'\'equation:
\begin{equation}\label{}
    ds^2 = \eta_{\mu\nu}dx^\mu dx^\nu = 0,
\end{equation}

laquelle relation est not\'ee par Kronheimer and Penrose (1967)
sous la forme suivante:
$$p \rightarrow q;$$

  \item La relation d'ant\'eriorit\'e temporelle (AT)
  (``chronological precedence'') d\'etermin\'ee par les droites
  de genre temps d'\'equation:
\begin{equation}\label{}
   ds^2  = \eta_{\mu\nu} dx^\mu dx^\nu > 0,
\end{equation}
not\'ee (Kron. and Pen., 1967):
$$ p \ll q;$$

  \item La relation d'ant\'eriorit\'e causale (``causal
precedence'') qui est la r\'eunion de ces deux relations (AH) et
(AT), not\'ee (Kron. and Pen., 1967):
\[
p \prec q.
\]
\end{enumerate}

Autrement dit, pour un espace-temps donn\'e $\mathcal{M}$ muni des
trois relations d'ant\'eriorit\'e ci-dessus $(
 \to , \ll , \prec)$, l'ensemble $
(\mathcal{M} \to , \ll , \prec )$ d\'efinit une structure causale
si les propositions suivantes sont satisfaites:

\begin{eqnarray*}
  \forall p \in \mathcal{M}, p \prec p,\\
  \forall (p,q,r) \in \mathcal{M}, (((p \prec q) \wedge (q \prec r)) \longrightarrow (p \prec r))
  ,\\
  \forall (p,q,r) \in \mathcal{M}, (((p \prec q) \wedge (q \prec p)) \longrightarrow (p = q))
  ,\\
  \forall p \in \mathcal{M}, \neg (p \ll p),\\
  \forall (p,q) \in \mathcal{M}, ((p \ll q) \longrightarrow (p \prec q)
  ),\\
  \forall (p,q,r) \in \mathcal{M}, (((p \prec q) \wedge (q \ll r))  \longrightarrow (p \ll r)
  ),\\
  \forall (p,q,r) \in \mathcal{M}, (((p \ll q) \wedge (q \prec r))  \longrightarrow (p \ll r))
  ,\\
  \forall (p,q) \in \mathcal{M}, ((p \to q) \Longleftrightarrow ((p \prec q) \wedge \neg (p \ll
  q)).\\
\end{eqnarray*}

Remarquons que, d'une part, la relation d'ant\'e\-riorit\'e causale
($\prec$) est localement une relation d'\textit{ordre
partiel}\footnote{Une relation d'ordre sur un ensemble $E$ est
dite partielle s'il existe au moins deux \'el\'ements de $E$
incomparables.}. D'autre part, la relation d'ant\'e\-rio\-rit\'e
horismotique ($\rightarrow$) caract\'erisant la structure du
c\^one isotrope est invariante sous le groupe des transformations
conformes\footnote{La transformation conforme du tenseur
m\'etrique $g $ est d\'efinie par la relation suivante: $ g_{\mu
\nu } \to \tilde g_{\mu \nu } = \Omega ^2 g_{\mu \nu } $ o\`u
$\Omega$ est une fonction r\'eelle, continue, non-singuli\`ere et
finie.} qui laissent invariant le tenseur m\'etrique. Par
cons\'equent, alors que dans le cadre newtonien l'ordre causal des
points-\'ev\'enements est isomorphe \`a l'ordre total d\'efini sur
l'ensemble des instants du temps absolu $t$ par la relation
d'ant\'eriorit\'e temporelle stricte $( \ll)$, sous cette
d\'efinition usuelle de la causalit\'e locale en relativit\'e,
l'ant\'eriorit\'e causale n'est pas identique \`a la seule
relation d'ant\'eriorit\'e temporelle.

Plus particuli\`erement, dans le cadre des transformations du
groupe de Lorentz $SO(1,3)$, le th\'eor\`eme de
Zeeman (1964) montre que si un \'ev\'enement $p$ est
ant\'erieur causalement \`a un \'ev\'enement $q$ dans un
r\'ef\'erentiel lorentzien $(R)$, il le sera \'egalement dans tout
autre r\'ef\'erentiel lorentzien $(R')$. Au contraire, pour des
\'ev\'enements $p$ et $q$ spatialement distincts qui seraient
newtoniennement simultan\'es dans le r\'ef\'erentiel $(R)$, l'on
pourrait trouver un r\'ef\'erentiel $(R')$ o\`u $p$ serait
ant\'erieur \`a $q$, et un autre o\`u $p$ serait post\'erieur \`a
$q$. Cette r\'egion des \'ev\'enements simultan\'es constitue le
domaine quadridimensionnel des points causalement non connectables
dans $(R)$, c'est-\`a-dire l'ailleurs spatial $ I^0$ relativement
\`a $(R)$ \`a un instant donn\'e.

La causalit\'e relativiste est donc localement dans la vari\'et\'e
espace-temps $(\mathcal{M},g)$ d'ordre spatio-temporel. Mais,
qu'en est-il de son ant\'erio\-ri\-t\'e? Elle semble toujours
contrainte par l'ant\'e\-rio\-ri\-t\'e chronologique 
(voir proposition 5 ci-dessus) 
bien que la proposition 8 exprime une contrainte de
l'ant\'eriorit\'e causale sans ant\'eriorit\'e temporelle. Nous
allons montrer dans la section suivante comment la construction
d'une topologie appropri\'ee afin de pr\'eserver la causalit\'e
usuelle (\textit{i.e.} la non-violation de l'ant\'eriorit\'e
temporelle dans $(\mathcal{M},g)$), nous permet de conclure que le
principe d'ant\'eriorit\'e sous-jacent \`a la relation causale
n'est pas r\'eductible \`a sa seule expression temporelle.

\section{Causalit\'e globale et topologie}
\subsection{Contraintes de courbure}

La structure causale est d\'efinie \textit{localement},
c'est-\`a-dire sur l'espace de Minkowski $T(\mathcal{M})$ tangent
en un point $p$ \`a la vari\'et\'e riemannienne espace-temps
$\mathcal{M}$, par le c\^one isotrope d'\'equation $ ds^2 = 0$.
Elle d\'etermine \textit{localement} un ordre partiel sur
$T(\mathcal{M})$ qui contraint les trajectoires physiquement
possibles \`a \^etre de genre temps (ou de genre isotrope) et
orient\'ees vers le futur.

L'\'equation d'une g\'eod\'esique de $\mathcal{M}$ pour un
syst\`eme de coordonn\'ees curvilignes quelconque $ \{ x^\mu \} $
est donn\'ee par:
\begin{equation}\label{}
    \frac{{d^2 x^\mu  }} {{ds^2 }} + \Gamma _{\nu
\lambda }^\mu  \frac{{dx^\nu  }} {{ds}}\frac{{dx^\lambda }} {{ds}}
= 0,
\end{equation}
o\`u $ {{d^2 x^\mu  } \mathord{\left/
 {\vphantom {{d^2 x^\mu  } {ds^2 }}} \right.
 \kern-\nulldelimiterspace} {ds^2 }}$ est la quadri-acc\'el\'eration de la particule dans un champ
gravitationnel. Le choix d'un r\'ef\'erentiel localement
lorentzien le long de la ligne d'univers $C$ de la particule
implique que les quantit\'es $ \Gamma _{\nu \lambda }^\mu $ sont
nulles dans la r\'egion infiniment petite qui entoure cette courbe
en tout point $p$ de $C$. De ce fait, on peut \'ecrire la relation
suivante valide localement:
\begin{equation}\label{}
    \frac{{d^2 x^\mu }} {{ds^2 }} = 0.
\end{equation}
La r\'eduction de l'intervalle \'el\'ementaire $$ ds^2  = g_{\mu
\nu } dx^\mu dx^\nu $$ \`a sa forme pseudo-euclidienne $$ ds^2  =
\eta _{\mu \nu }dx^\mu dx^\nu   = diag( +1, - 1, - 1, - 1) $$
permet de construire en chaque point $p$ de $C$ un c\^one isotrope
ayant ce point $p$ comme sommet. Ici, la trajectoire d'une
particule libre dans l'espace-temps courbe $\mathcal{M}$ est la
donn\'ee d'une courbe $C$ telle que la quadri-vitesse unitaire $
{{dx^\mu  (\xi )} \mathord{\left/
 {\vphantom {{dx^\mu  (\xi )} {d\xi }}} \right.
 \kern-\nulldelimiterspace} {d\xi }}$, tangent en $p$ \`a $C$, est confin\'ee \`a l'int\'erieur de tout
c\^one isotrope dont le sommet $p$ est situ\'e sur elle. Mais,
alors que $C$ est une g\'eod\'esique de $\mathcal{M}$, sa
trajectoire correspondante dans l'espace de Minkowski
$T(\mathcal{M})$, tangent en $p$ \`a $\mathcal{M}$, est une
trajectoire curviligne qui fait intervenir les principes de la
dynamique. N\'eanmoins, l'\'equivalence locale, au niveau des
connexions, entre la vari\'et\'e riemannienne et l'espace
pseudo-euclidien tangent ne peut se maintenir globalement au
niveau des courbures (Synge 1960, p. IX.) Les quantit\'es $
g_{\mu\nu}$, qui d\'eterminent le champ de gravitation au
voisinage de tout point $p$ de $C$ dans $(\mathcal{M},g)$,
d\'eforment localement les c\^ones isotropes contenant $C$.
Autrement dit, l'ant\'eriorit\'e causale devra \^etre assur\'ee
dans $\mathcal{M}$ par des \textit{conditions topologiques}
appropri\'ees.

\subsection{Conditions topologiques}
Dans un espace-temps plat de Minkowski $(\mathcal{M},\eta)$, la
relation causale d\'efinit une topologie d'Alexandroff\footnote{La
donn\'ee d'une topologie d'Alexandroff est \'equivalente \`a la
donn\'ee d'une structure d'ordre partiel, c'est-\`a-dire d'une
relation r\'eflexive, transitive et antisym\'etrique. Or, les
relations d'ant\'eriorit\'e temporelle et causales d\'eterminent
sur $\mathcal{M}$ un ordre partiel. Autrement dit, ``Essentially,
it identifies the basis of open sets of the topology with sets of
events timelike accessible from a pair of events, i.e., an open
set in the basis is the common region of the interior of a forward
light cone from one event and the interior of the backward light
cone from another'' Sklar 1985, p. 255.} qui co\"{i}ncide avec la
vari\'et\'e topologique usuelle, ou espace de
Hausdorff\footnote{Un espace topologique est dit de Hausdorff en
tant qu'il v\'erifie l'axiome $(T_{2})$ de s\'eparation, \`a
savoir, quels que soient deux points diff\'erents $p$ et $q$ dans
$\mathcal{M}$, il existe un voisinage de $p$ et un voisinage de
$q$ sans point commun.}:

\begin{quote}
``In the Minkowski spacetime of special relativity we can indeed
causally define [...] the open set basis sufficient to fully
define the topology. An explicit definition of open sets in terms
of causal connectibility is available in terms of the well-known
Alexandroff topology for Minkowski spacetime''. (Sklar 1985)
\end{quote}

En effet, $\forall (p,q) \in M | p \ll q, $ et pour tout point $r
\in \mathcal{M}$, l'ensemble $ {\{ r | p < r < q\}} $ d\'efinit
une topologie d'Alexandroff sur $\mathcal{M}$. Cependant, ce
r\'esultat ne serait \^etre \'etendu dans un espace-temps
pseudo-riemannien \`a moins de munir celui-ci d'une hi\'erarchie
de conditions de causalit\'e (Hawking 1974). Si l'on
souhaite, en effet, conserver la structure causale qu'implique le
groupe de Lorentz comme un invariant dans toute transformation de
coordonn\'ees, la vari\'et\'e riemannienne $\mathcal{M}$ doit
\^etre un construit topologique appropri\'e. Comme le souligne Joshi (1996):

\begin{quote}
``The local causality principle for a space-time implies that over
a small regions of space and time the causal structure is the same
as in the special relativity. However, as soon as one leaves the
local domain, global pathological features may show up in the
space-time such as the violation of time orientation, possible
non-Hausdorff nature or non-paracompactness, having disconnected
components of space-time, and so on. Such pathologies are to be
ruled out by means of `reasonable' topological assumptions only''.
\end{quote}

Montrons comment est construite cette topologie appropri\'ee. Soit
une vari\'et\'e espace-temps $\mathcal{M}$. Elle satisfait \`a la
relation d'ant\'eriorit\'e causale (AH) est impl\'ement\'ee des
conditions d'impossibilit\'es suivantes:

\begin{description}
  \item[(1)] \noindent  Une courbe ferm\'ee $\gamma $ de genre temps
  est impossible dans $\mathcal{M}$, \`a savoir: $$\forall p
\in \mathcal{M}, p \notin I^ + (p) .$$ De la sorte, le futur
chronologique de $p$ est l'ouvert contenant tous les points $q$
tel que $$ I^ + (p) \equiv \left\{ {q \in \mathcal{M} | p \ll q}
\right\},$$ et, respectivement, le pass\'e chronologique de $p$
est l'ouvert contenant tous les points $q$ tel que $$ I^ -  (p)
\equiv \left\{ {q \in \mathcal{M} | q \ll p} \right\}.$$

  \item[(2)] \noindent Une courbe ferm\'ee $ \gamma$ de genre temps ou
   isotrope est impossible dans $\mathcal{M}$, \`a savoir:
$$\forall p \in \mathcal{M}, p \notin I^ +  (p) \vee \dot J^ +
(p).$$ Le futur causal de $p$ est l'ouvert contenant tous les
points $q$ tel que:
$$ J^ +  (p) \equiv \left\{ {q \in \mathcal{M} | p \prec q}
\right\},$$ et, respectivement, le pass\'e causal de $p$ est
l'ouvert contenant tous les points $q$ tel que: $$ J^ - (p) \equiv
\left\{ {q \in \mathcal{M} | q \prec p} \right\}.$$
\end{description}


De ces conditions d'impossibilit\'e, il en r\'esulte qu'une
vari\'et\'e $\mathcal{M}$ est causale au sens usuel s'il est
impossible d'identifier topologiquement des points de
$\mathcal{M}$ dans le temps\footnote{``[...] In physically
realistic solutions, the causality and chronology conditions are
equivalent'' (Hawking 1974, \S6.4.5}. En outre, trois autres
conditions topologiques sont aussi n\'ecessaires pour interdire
une structure globale o\`u un point d'une courbe causale $ \gamma$
pourrait arbitrairement avoisiner soit le point d'origine de $
\gamma$, soit une autre courbe causale proche de ce m\^eme point
d'origine, \`a savoir:

\begin{enumerate}
  \item La topologie prescrite sur $\mathcal{M}$ est d'Alexan\-droff
(Penrose 1972), ou bien la topologie d'Alex\-an\-droff est de
Hausdorff;

  \item La topologie prescrite sur $\mathcal{M}$ est
non-compacte\footnote{``If M is chronological, M cannot be
compact" (Joshi 1996)} (voir Esposito 1992);

  \item La vari\'et\'e est globalement hyperbolique\footnote{``Let $M$ be
globally hyperbolic, then $M$ is homeomorphic to $\mathbb{R}\times
S$ where $S$ is a three-dimensional submanifold and for each $ t
\in \mathbb{R}$, $\{t\}\times S$ is a Cauchy surface for $M$" 
(Joshi 1996)}, c'est-\`a-dire l'ensemble ${J^ -  (q) \cap J^ +
(p)}$ est compact. En d'autres termes, la vari\'et\'e
$\mathcal{M}$ admet une foliation globale de type:
$$\mathcal{M} = \Sigma _{t} \times \mathbb{R},$$ avec $\Sigma _{t}$
une hypersurface de Cauchy (voir aussi Barrow 1988, \S10.3).

\end{enumerate}

\subsection{Conditions de causalit\'e}
A ces conditions topologiques correspond une hi\'erarchie de
conditions causales (Carter 1971) ordonn\'ees par ordre de
restriction croissante:

\begin{enumerate}
  \item \textit{La condition de causalit\'e forte}. La vari\'et\'e $\mathcal{M} $
est dite distingu\'ee dans le temps (Kron. and Pen., 1967) selon la propri\'et\'e suivante:\[
\forall (p,q) \in \mathcal{M}, ((I^ +  (p) \equiv I^ +  (q))
\longrightarrow (p = q)).
\] La topologie induite par la structure causale co\"{i}ncide alors avec
la vari\'et\'e topologique usuelle\footnote{``Let $(M,g)$ be a
strongly causal space-time. Then the manifold topology on $M$ is
the same as the Alexandroff topology" (Joshi 1996)}. En d'autres
termes, ce sont les relations d'ant\'eriorit\'e temporelle qui
d\'eterminent la topologie de la vari\'et\'e espace-temps. De
plus, Carter (1971) a montr\'e qu'il existe une hi\'erarchie
\textit{non-d\'enombrable} de conditions causales  de plus en plus
restrictives, de telle sorte que des courbes causales presque
ferm\'ees sur elle-m\^emes sont interdites dans $\mathcal{M}$.

  \item \textit{La condition de causalit\'e stable}. Il existe en tout point $p$ de
$\mathcal{M} $ une fonction de temps global, i.e. un champ
scalaire lisse $f$ dont le gradient est partout de genre temps
(Hawking 1974, \S6.4.9).

La condition de stabilit\'e causale introduit une transitivit\'e
de la relation de simultan\'eit\'e \`a l'\'echelle globale 
(Kerszberg 1989, 1994). Mais un temps universel
n'est pas n\'ecessairement le temps absolu de la physique newtonienne

\begin{quote}``qui est, dit Newton (1687), sans relation
\`a quoi que ce soit d'ext\'e\-rieur, en lui-m\^eme et de par sa
nature coule uniform\'ement''.
\end{quote}

Ce qui caract\'erise un temps universel, c'est qu'il se laisse
construire g\'eom\'etriquement \`a partir de la transitivit\'e
globale de la relation usuelle de simultan\'eit\'e, celle
d\'efinie par des 3-plans de genre espace orthogonaux aux lignes
d'univers des observateurs galil\'eens (Cartan  1923 et une
synth\`ese par Ruede et Straumann 1997).

Il en est tout autrement quand il s'agit de la relativit\'{e}, qui
remplace la structure d'espace fibr\'e $(M,h,\tau )$ de l'espace-temps
newtonien par une structure
m\'etrique d'espace-temps unique $(\mathcal{M},g)$. Mais, on peut
passer au cas limite de la physique newtonienne en posant
localement
\[ c \to \infty,
\]
ce qui fournit le 3-plan newtonien de genre espace auquel se
r\'eduit la r\'egion de l'ailleurs spatial $ I^0$ lorsque les
nappes $ \Gamma^ +$ et $ \Gamma^ -$ du c\^one isotrope tendent
l'une vers l'autre. Il convient de remarquer que par analogie avec
cette notion usuelle de la simultan\'eit\'e on peut d\'efinir dans
un r\'ef\'erentiel $R$ de coordonn\'ees lorentziennes une surface
de genre espace $ S$ dont chaque point est situ\'e dans l'ailleurs
$ I^0$ de tout autre point, de telle sorte qu'il existe un
ensemble d'\'ev\'enements simultan\'es relativement \`a $ S$.

Dans ce cadre, un observateur $R$ peut op\'erer une
foliation\,\footnote{En r\'ealit\'e, pr\'ecisons qu'il n'existe
pas \textit{a priori} dans $ (\mathcal{M}, \eta)$ de telle
foliation par des surfaces de genre espace.} de la vari\'et\'e $
\mathcal{M}$ par une famille d'hyperplans de genre espace, sans
intersection et normaux \`a sa ligne d'univers en chacun de ses
instants de temps propre. La vari\'et\'e espace-temps
$\mathcal{M}$ peut alors \^etre d\'ecompos\'ee en une partie
temporelle $ \mathbb{R}$ et une partie purement spatiale $ S$
(l'espace-quotient):
\[ \mathcal{M} = S \times \mathbb{R}.
\]
Comme chaque hypersurface de genre espace $ S$ est relative \`a un
observateur donn\'e, il existe (Geroch 1981) autant de
foliations de $\mathcal{M}$ que d'observateurs en chute libre.
Cela signifie que pour ces observateurs, ni l'indication du temps
ni l'indication d'espace n'ont de valeur absolue ; l'une et
l'autre d\'ependent de leur \'etat de mouvement. 

  \item \textit{La condition de d\'eterminisme causal}, au sens o\`u la
vari\'et\'e $\mathcal{M}$ est un espace-temps physiquement
pr\'edictible. Dans une vari\'et\'e $\mathcal{M}$ globalement
hyperbolique il existe une fonction de temps, dit de Cauchy, telle
que les donn\'ees initiales sur une hypersurface de genre espace
suffisent \`a d\'eterminer les propri\'et\'es de la vari\'et\'e
$\mathcal{M}$ dans son ensemble\footnote{Selon les mots de S.
Chandrasekhar (1987): ``The causal character of the laws of physics
requires that, given complete initial data on a space-like
three-surface, the future is uniquely determined in the space-time
domain bounded by the future-directed in-going null rays emanating
from the boundary of the spatial slice''.}.

Donnons-nous $\mathcal{M}$, comme construit topolo\-gi\-que de
l'ensemble des points dans l'espace tri-dimensionel $S$ \`a un
instant $t$ de temps cosmique donn\'e. Par d\'efinition,
$\mathcal{M}$ est globalement hyperbolique si les ensembles $ {J^
+ (q) \cap J^ - (p)}$ sont des compacts pour tout point $(p, q)
\in \mathcal{M}$.

Pour un ouvert $S \subset \mathcal{M}$, on d\'efinit \'egalement:

\begin{itemize}
  \item $\forall p \in \mathcal{M}, J^ +  (S) \equiv \bigcup\limits_{p \in S} {J^ +  (p)},$
  \item $\forall p \in \mathcal{M}, J^ -  (S) \equiv \bigcup\limits_{p \in S} {J^ -  (p)}.$
\end{itemize}

L'ensemble $S$ est une surface de Cauchy $ \Sigma _t$ si nous
avons les propri\'et\'es topologiques suivantes:

\begin{itemize}
  \item $S$ est un ferm\'e, non-vide et achronal, tel que: $$ \forall (p,q) \in \mathcal{M}, ((q \in
  S)
\rightarrow  (q \notin I^ +  (p))) ,$$ \noindent c'est-\`a-dire $
I^ + (S) \cap S \equiv \emptyset $;
\bigskip
  \item $S$ est une feuille de $\mathcal{M}$, c'est-\`a-dire une
hypersurface achronale sans fronti\`ere topologique, not\'ee $
\dot S$\footnote{La fronti\`ere topologique est l'ensemble des
points $p$
  tel que dans tout voisinage $\mathcal{U}$ de $p$ il existe les
  points $ q \in I^ +  (p)$ et $ r \in I^ -  (p)$ et une courbe du
  genre temps passant par $q$ et $r$ qui ne coupe pas $S$ (Wald 1984, \S8.3}.
\bigskip
 \item $\mathcal{D}^{+}(S) $ est l'ensemble des points-\'ev\'ene\-ments
de $\mathcal{M}$ reli\'es \`a $p$ par des lignes causales du
pass\'e coupant $S$ une seule fois;
\bigskip
  \item $\mathcal{D}^{-}(S)$ est l'ensemble des points-\'ev\'ene\-ments de $\mathcal{M}$
reli\'es \`a $p$ par des lignes causales du futur coupant $S$ une
seule fois.
\end{itemize}

Le domaine de d\'ependance $\mathcal{D}(S)$ est l'union de
$\mathcal{D}^{+}(S)$ et de $\mathcal{D}^{-}(S)$. La condition
$\mathcal{D}^{+}(\Sigma _{t})$ est plus restrictive que les
conditions d'ant\'eriorit\'e causale ou temporelle, \`a savoir:
$J^ + (p) $ ou $I^ + (p)$ (et respectivement $\mathcal{D}^{-}(S)$
par rapport \`a $J^ - (p) $ ou $I^ -  (p)$); car elle ajoute une
contrainte de pr\'edictibilit\'e. En effet, si $p$ $\in$
$\mathcal{D}^{+}(S)$\footnote{Si $p$ et $q$ $\in \Sigma _{t}$,
alors $q \notin I^ +  (p), I^ +  (\Sigma _{t} ) \cap \Sigma _{t}
\equiv \emptyset$.} alors $p \in J^ + (S)$ est compl\`etement et
univoquement d\'etermin\'e. Dans une vari\'et\'e $\mathcal{M}$
globalement hyperbolique o\`u $\mathcal{M} = D(S)$, la totalit\'e
de l'espace-temps est donc d\'eterminable \`a partir des donn\'ees
obtenues sur une seule hypersurface achronale de Cauchy $\Sigma
_{t}$.
\end{enumerate}

En relativit\'e g\'en\'erale, l'ant\'eriorit\'e causale est donc
un construit chrono-g\'eom\'etrique dont la propri\'et\'e usuelle
d'ant\'eriorit\'e temporelle stipule une topologie appropri\'ee.
En effet, une des cons\'equences significatives de cette fa\c{c}on
d'imposer une hi\'erarchie de conditions causales \`a la
g\'eom\'etrie de l'espace-temps, est que la r\'ef\'erence \`a un
temps cosmique, ``comme newtonien'', sous-jacente \`a
l'ant\'eriorit\'e causale prise dans son sens usuel, doit \^etre
associ\'ee \`a des propri\'et\'es topologiques particuli\`eres de
la vari\'et\'e espace-temps permettant de distinguer globalement
dans $\mathcal{M}$ des feuilles d'espace simultan\'e $\Sigma_{t}$.
Autrement dit, l'ant\'eriorit\'e causale est elle-m\^eme
spatio-temporelle.

\section{Conclusion: niveaux d'ant\'eriorit\'e causale\ \ \ }

Le cadre conceptuel de la causalit\'e auquel nous a habitu\'e la
physique depuis Newton est marqu\'e par la construction
diff\'erentielle d'une relation d'ant\'ec\'edent \`a cons\'equent
selon, \`a titre principal, un principe d'ant\'eriorit\'e
temporelle. Par cons\'equent, la causalit\'e physique est
usuellement caract\'eris\'ee par une notion d'ant\'eriorit\'e
simplement de ``souche temporelle'' extrins\`eque aussi bien \`a
la m\'etrique qu'\`a la topologie de l'espace o\`u se d\'eroule
les ph\'enom\`enes naturels. Mais, suite aux d\'eveloppements du
concept d'espace-temps relativiste, la r\'ef\'erence au temps dans
la relation d'ant\'eriorit\'e causale ne peut plus \^etre
analys\'ee sans forme math\'ematique, dont la structure
chrono-g\'eom\'etrique d\'epend des ph\'enom\`enes dynamiques
qu'elle organise. Puis, la construction des diverses conditions de
causalit\'e permet d'\'eclaicir les liens qui existent entre les
niveaux de causalit\'e (causalit\'e locale et causalit\'e globale)
et notamment entre l'existence d'un temps cosmique et la topologie
de la vari\'et\'e espace-temps.

Nous proposons d'avancer trois niveaux distincts d'ant\'eriorit\'e
causale selon trois niveaux d'ant\'eriorit\'e temporelle:

\begin{enumerate}
\item \textbf{\textit{Premier niveau (ordre descriptif)}}:

La relation causale satisfait \`a la limitation du c\^one
isotrope, en tant que la forme diff\'erentielle quadratique
$g_{\mu\nu} $ est r\'eductible localement \`a la m\'etrique
lorentzienne $\eta_{\mu\nu}$ (th\'eor\`eme de platitude locale).
La condition de premier niveau est une condition de
\textit{lin\'earisation} de la structure causale de $
\mathcal{M}_{4}$ par le passage \`a l'infiniment petit. Elle
affirme que localement la loi de causalit\'e est
\textit{d\'ecrite} par la causalit\'e usuelle. Car, on peut
toujours trouver dans une r\'egion infint\'esimale de $
\mathcal{M}$ un syst\`eme de coordonn\'ees lorentziennes (section
2);

\item \textbf{\textit{Deuxi\`eme niveau (ordre constructif)}}:

La relation causale satisfait aux conditions de Carter (ou
conditions dites de ``raisonabilit\'{e}
physique''(Heller  1990), en tant que la distinction entre le
temps et l'espace est une propri\'et\'e globale de l'espace-temps.
Ce second niveau affirme que le tenseur de courbure est totalement
d\'etermin\'e par son tenseur de courbure spatiale. On peut
construire une coordonn\'ee de temps de fa\c{c}on qu'\`a tout
instant la m\'etrique soit partout la m\^eme dans tout l'espace.
En d'autres termes, la condition de second niveau consiste \`a
supposer fixe, parmi les \^etres du groupe plus \'etendu des
transformations continues des coordonn\'ees, $Diff(\mathcal{M}) $,
le c\^one isotrope comme un \^etre sp\'ecial ou ``absolu'', dans
le nouvel espace-temps courbe de la relativit\'e g\'en\'erale,
parce qu'il ``limite'' localement la causalit\'e (section 3).
L'ant\'eriorit\'e temporelle d\'epend ici d'une prise
\textit{constructive} sur toutes les topologies possibles, dont le
sch\'ema doit satisfaire d'une mani\`ere explicite \`a notre
id\'ee usuelle du temps newtonien;

\item \textbf{\textit{Troisi\`eme niveau (ordre constitutif)}}:

La relation causale satisfait \`a l'ensemble des solutions des
\'equations d'Einstein. Si les mod\`eles d'espace-temps sans
l'existence d'un temps cosmique ne sont pas physiquement possibles
(plut\^ot qu'admissibles), c'est qu'il existerait des contraintes
dynamiques ``ant\'erieures'' \`a la contrainte d'ant\'eriorit\'e
spatio-temporelle (section 3). La causalit\'e d\'epend toujours
d'une prise constructive sur toutes les topologies possibles, mais
qui doit aussi satisfaire d'une mani\`ere explicite aux
comportements permis par le syst\`eme dynamique dont l'\'evolution
est d\'ecrite par le syst\`eme des \'equations du champ
d'Einstein. Le troisi\`eme niveau, plus conjectural, signifie une
analyse conjointe  de la g\'eom\'etrie de l'espace-temps et des
syst\`emes dynamiques. Cette analyse fait l'objet d'un article \`a
venir.
\end{enumerate}


\end{document}